\documentclass[twoside]{article}

\usepackage{amsmath, amssymb}
\usepackage{graphicx,psfrag,epsf}
\usepackage{enumerate}
\usepackage{url} 
\usepackage{dcolumn}
\usepackage{fullpage}
\usepackage{mathtools}

\usepackage{algorithmicx, algorithm,algpseudocode}
\usepackage{tikz}
\usetikzlibrary{fit,positioning}

\usepackage{fancyhdr}

\newcommand\YS{YS}
\newcommand\G{\Gamma}
\newcommand\E{\mathbb{E}}
\newcommand{\lambdaIter}[1] {\lambda^{(#1)}}
\newcommand{\indices}[1]{i=1,\dots,#1}
\newcommand{\lambdaHatEM} {\hat{\lambda}^{(EM)} }
\newcommand{\lambdaHatM} {\hat{\lambda}^{(M)} }

\graphicspath{{/Users/rdenisa/Documents/research/yule/figures}{/Users/rlucas/Documents/ResearchPapers/notes_yule_simon/ai_stats/figures/}}

\begin{document}


%

%

\twocolumn[

\author{
  Roberts, Lucas\\
  \texttt{rlucas@amazon.com}
  \and
  Roberts, Denisa\\
  \texttt{rdenisa@amazon.com}
}

\title{An Expectation Maximization Framework for Yule-Simon Preferential Attachment Models}
\maketitle
]

%



\begin{abstract}
In this paper we develop an Expectation Maximization(EM) algorithm to estimate the parameter of a Yule-Simon distribution. 
The Yule-Simon distribution exhibits the ``rich get richer'' effect whereby an 80-20 type of rule tends to dominate. 
These distributions are ubiquitous in industrial settings. 
The EM algorithm presented provides both frequentist and Bayesian estimates of $\lambda$. 
By placing the estimation method within the EM framework we are able to derive Standard errors of the resulting estimate. 
Additionally, we prove convergence of the Yule-Simon EM algorithm and study the rate of convergence. An explicit, closed form solution for the rate of convergence of the algorithm is given.  
\end{abstract}

\section{INTRODUCTION}

The expression ``the rich get richer'' in the machine learning community is also known as a Preferential Attachment (PA) model. The effect occurs because wealthy individuals are more likely to have funds to invest in future profitable business ventures, thereby increasing their wealth. Furthermore, the rich are exposed to more investment opportunities and thereby have more chances to make profit maximizing investment decisions. However, this effect is not limited to business endeavors, in models of text generation the preferential attachment model assumes that an author who has already used a word in a text is more likely to use the word again \cite{simon1955class}. In the text generation model, the ``richness'' of the word is how many times the author uses the word in the text. In a PA model the frequency of an item is inversely proportional to the rank of the item raised to a non-negative power $\lambda$. There are many other applications of the PA model such as the number of articles published by authors \cite{price1976general,simon1955class}, network models\cite{barabasi1999emergence}, and evolution of taxa \cite{yule1925mathematical,simon1955class}. A preferential attachment process is one where a new item will be added to an existing category with probability $\alpha$ and a new category will be added to the set of existing categories with probability $1-\alpha$. Each new arrival is more likely to belong to the most popular existing category. Yule \cite{yule1925mathematical} and Simon \cite{simon1955class} each derived a formal model for this verbal description of the stochastic process. 
Garcia proposed a fixed point algorithm to get a maximum likelihood estimator for the parameter of the Yule-Simon (YS) distribution \cite{garcia2011fixed}. Alternatives to the maximum likelihood estimation has been employed across other application domains such as median ranked regression in reliability engineering \cite{olteanu2008technical}.
Furthermore, in random graph theory, when studying the degree distribution of random networks often a least squares algorithm of the log-log degree distribution of the nodes provides a least-squares estimate of the parameter $\lambda$ \cite{bak2013nature}. 

Previous work by Leisen et al. \cite{leisen2016objective,leisen2017note} used the mixture representation of the \YS\ to derive a Gibbs sampler for the posterior distribution of $\lambda$. The Gibbs algorithm requires hyperparameter selection, burn in, and thinning selections in addition to sampling many times from the posterior. While Gibbs sampling can converge under certain conditions, to our knowledge this remains to be shown for the \YS\  distribution. In contrast the EM algorithm has convergence guarantees and convergence rate quantifications \cite{balakrishnan2017statistical, ma2000asymptotic,xu1997comparative,dempster1977maximum}. Additionally, the EM algorithm is reproducible and is not complicated by stochasticity considerations like the Gibbs sampler. 


This article's contributions are: (1) an EM algorithm to estimate the parameter of the \YS\ distribution, (2)  standard errors of the estimate using the Louis \cite{louis1982finding} and Oakes \cite{oakes1999direct} methods, (3) a proof of convergence of the EM algorithm for the \YS\ parameter $\lambda$, and (4) estimation of the convergence rate of the EM updates.  Additionally we present extensive experiments to empirically validate our equations and derivations. We compare our estimates to both those provided by the fixed point algorithm and to the estimates of the Gibbs sampler of the posterior distribution for $\lambda$. We provide an empirical application to real text using five novels from project Gutenberg \cite{projGut}. 

\section{THE \YS\ MODEL}\label{sec:ys_model}

In this section we describe formally the \YS\ model and define our notation that will be used throughout the remainder of the manuscript. 

\subsection{A Mixture Representation of the \YS\ Model}


Let $\lambda$ denote the rate parameter for the \YS\ distribution of the random variable $K$, where $k_i$ are the observed counts of the $i$th observation from the Yule-Simon process, with $\indices{N}$. Then the probability mass function may be written
\begin{equation}\label{eqn:dens}
g(k_i\vert \lambda) = \lambda\textnormal{B}(\lambda+1,k_i), 
\end{equation}
where \textnormal{B}$(\cdot, \cdot)$ denotes the beta special function. Equivalently
\begin{equation}\label{eqn:ys_w_gammas}
g(k_i \vert \lambda) = \lambda\frac{\Gamma(k_i)\Gamma(\lambda + 1)}{\Gamma(k_i + \lambda + 1)},
\end{equation}
where $\Gamma(x) = \int_0^\infty z^x e^{-z}dz$, the gamma function. A mixture representation for the \YS\ distribution was given by Leisen et al. \cite{leisen2017note} and earlier by Devroye\cite{devroye:1986}. If $k_i$, $\indices{N}$, arise from a \YS\ process, then the $k_i$ may be represented by the following stochastic model

\begin{equation}\label{eqn:latents}
\begin{aligned}
w_i \sim&\ \textnormal{Exponential}(\lambda).\\
p_i =&\ e^{-w_i}, \\
k_i \sim&\ \textnormal{Geometric}(p_i), \\
\end{aligned}
\end{equation}

Here $P$ are latent random variables that define the success probability in the geometric Bernoulli trials.
Therefore the joint distribution of $k_i$ and $p_i$, $\indices{N}$, is 
\begin{equation}\label{eqn:joint}
\begin{split}
f(k_i, p_i \vert \lambda) &= p_i^{\lambda}(1-p_i)^{(k_i-1)}\frac{\G(\lambda+1)}{\G(\lambda)}\\
&= \lambda p_i ^{\lambda}(1-p_i)^{(k_i - 1)}.
\end{split}
\end{equation}
Equation \ref{eqn:joint} in EM applications is referred to as the \emph{complete data likelihood}. Integrating over $p_i \in (0,1)$ returns Equation \ref{eqn:dens}. We interpret this mixture representation as a missing data problem, the Bernoulli probability is not measured at each trial. In fact these Bernoulli probabilities may be interpreted as novelty seeking probabilities. 

Additionally, in the application of the EM algorithm, we require the distribution of the latent or unobserved data given the observed data. 
The missing data distribution is a distribution over the geometric probability $p_i$ given the observed data $k_i$. Now proceeding via Bayes' rule and using Equation \ref{eqn:dens} and Equation \ref{eqn:joint} to compute the conditional distribution of $p_i$ given $k_i$, $\indices{N}$,
\begin{equation}\label{eqn:condit}
\begin{split}
h(p_i\vert k_i, \lambda) &= p_i^\lambda(1-p_i)^{k_i-1}\frac{\G(\lambda+1+k_i)}{\G(\lambda+1)\G(k_i)}\\
&= Beta(\lambda + 1, k_i).
\end{split}
\end{equation}
To build intuition with the \YS\ distribution it is useful to consider some generic properties of the mass function; the probabilities are strictly decreasing; the modal count is always one; the mean of the random variable is $\lambda /(\lambda-1)$ for $\lambda >1$, otherwise the mean is infinite; the variance does not exist when $\lambda <2$.
\section{Estimates and Standard Errors}
Next we present two estimation methods for the \YS\ distribution. 
\subsection{A Gibbs Algorithm for the Posterior of the \YS\ Parameter}
In this section we contrast the use of the EM algorithm to estimate $\lambda$ with Gibbs sampling of the posterior for $\lambda$.
Gibbs sampling from the posterior of a \YS\ PA process is derived by Leisen et al. \cite{leisen2017note}. The main interest is to numerically contrast both the $\lambda$ estimates and the standard errors between the two methods. 
Conversely, Leisen et al. contend that Gibbs exploits the Bayesian property of good performance in small samples.
To sample from the posterior $\lambda \vert \bf{w},\bf{k}$ we use the algorithm from Leisen et al. given by the following probability sampling process 
\begin{itemize}
\item Sample $p_i\vert \lambda$, $k_i$ ~ $Beta(\lambda + 1, k_i)$,
\item Compute $w_i = -log(p_i)$, $i=1,\dots,N$,
\item Sample $\lambda\vert \bf{w},\bf{k}$ ~ $Gamma(a+n, b = \sum_{i=1}^N w_i)$,
\end{itemize}
where $a$ and $b$ are the shapes of a Gamma prior, $k_i$ are the observed data and $w_i$ are latent variables in the formulation expressed in Equation \ref{eqn:latents}. In this paper all logarithms are natural or base $e$ and bold letters ($\bf{w}$) denote vectors composed of elements $w_i$. 

\begin{figure}
\centering
\begin{tikzpicture}
\tikzstyle{main}=[circle, minimum size = 0mm, thick, draw =black!80, node distance = 16mm]
\tikzstyle{connect}=[-latex, thick]
\tikzstyle{box}=[rectangle, draw=black!100]
  \node[main] (w) [ circle,label=below:w] {};
  \node[main] (beta) [left=of w,label=below:$\lambda$] { };
  \node[main, fill = black!10] (k) [right=of w,label=below:k] { };
  \node[main] (a) [left=of beta,circle,fill,inner sep=1pt,label=below:a] {};
  \node[main] (b) [above=of beta,circle,fill,inner sep=1pt,label=right:b] {};
  \path (w) edge [connect] (k)
		(beta) edge [connect] (w);
  \path(a) edge [connect] (beta);
  \path(b) edge [connect] (beta); 
    \node[rectangle, inner sep=0mm, fit= (w) (k),label=above left:N, xshift=19mm] {};
    \node[rectangle, inner sep=4.4mm,draw=black!100, fit= (w) (k)] {};
\end{tikzpicture}

\caption{A graphical model depiction of the mixture representation of the model.}

\label{fig:dag_pic}
\end{figure}
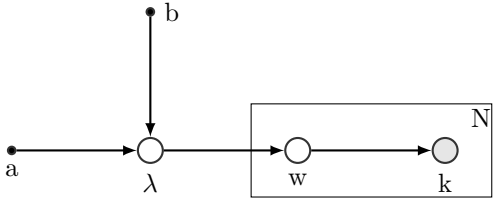

A graphical model representation is depicted in Figure \ref{fig:dag_pic}.

\subsection{EM Algorithm}
In this section we derive an EM algorithm to estimate the rate parameter of the \YS\ distribution. 
First consider the complete data log-likelihood  
\begin{equation}
l(\lambda; k_i, p_i)= \sum_{i=1}^N(k_i-1) log(1-p_i) + \lambda log(p_i) + log(\lambda),
\end{equation}
for $\indices{N}$. Now considering a single observation $i$, we form the $Q_i(\cdot \vert \cdot)$ function which is the expectation of the complete data likelihood where the expectation is with respect to the density in Equation \ref{eqn:condit}. 
Then 
\begin{equation}\label{eqn:qi}
\begin{aligned}
Q_i(\lambda \vert \lambdaIter{t}) 
&= \lambda \E[log(p_i)] + (k_i-1)\E[log(1-p_i)]  \\
& \ \ + log(\lambda). 
\end{aligned}
\end{equation}
Note the value of the parameter at the previous iteration, denoted $\lambdaIter{t}$, enters through the conditional expectations because the expectations are under $h(p_i\vert k_i, \lambdaIter{t})$ given in Equation \ref{eqn:condit}. 
The expectation in Equation \ref{eqn:qi} are entropies of beta distributions given by differences of digamma functions. Derivations and equations are given in the appendix.  
Finally, we sum over all values of $\indices{N}$ and simplify the Q-function to
\begin{equation}\label{eqn:q_func}
\begin{split}
Q(\lambda \vert \lambdaIter{t}) &= N\lambda\psi( \lambdaIter{t} +1) - \lambda\sum_{i=1}^N\psi( \lambdaIter{t} +1 + k_i) + \\
& \sum_{i=1}^N(k_i-1)\left[\psi(k_i) - \psi( \lambdaIter{t} +1 + k_i)   \right] \\
&+ N\log(\lambda).
\end{split}
 \end{equation}
To determine the updating equation for $\lambda$ we take a derivative of $Q$ with respect to $\lambda$ and calculate
 \begin{equation}\label{eqn:partial_q}
\frac{\partial Q}{\partial \lambda} = N\psi( \lambdaIter{t} +1) - \sum_{i=1}^N\psi( \lambdaIter{t} +1 + k_i) +\frac{N}{\lambda}.
 \end{equation}
Setting Equation \ref{eqn:partial_q} equal to 0 we determine
 \begin{equation}\label{eqn:fixed_psi}
 \psi( \lambdaIter{t} +1) +  \frac{1}{\lambda} =  \frac{1}{N}\sum_{i=1}^N\psi( \lambdaIter{t} +1 + k_i). 
 \end{equation}
Equation \ref{eqn:fixed_psi} is also the equation solved by the fixed point algorithm in Garcia \cite{garcia2011fixed}. 
The EM algorithm may be interpreted as a gradient descent method on the observed data likelihood. 
The EM estimate is the fixed point of the \emph{self consistency equation} \cite{flury2000exercises}
\begin{equation}\label{eqn:em_used}
\lambdaIter{t+1} = \frac{N}{\sum_{i=1}^N (\psi( \lambdaIter{t} +1 + k_i) - \psi( \lambdaIter{t} +1) )}. 
\end{equation}
Similarly to Garcia we leverage the digamma function recursion given by 
 \begin{equation}\label{eqn:digamma}
 \psi(z+1) = \psi(z) + \frac{1}{z}, 
 \end{equation}
and iterating this recursion gives us a finite sum representation to get the $\lambda$ updates at each EM iteration,
 \begin{equation}\label{eqn:em_update_eqn}
\lambdaIter{t+1} =   \frac{N}{ \sum_{i=1}^N\sum_{j=1}^{k_i} \frac{1}{\lambdaIter{t}+j} }. 
 \end{equation}
 
The EM algorithm also can estimate a \emph{maximum a posteriori} estimate inside a Bayesian framework. The algorithm adds a $log(\pi(\lambda))$ term to the $Q(\cdot \vert \cdot)$, where $\pi(\lambda)$ is the prior distribution. Following the prior choice in Leisen et al. we choose a gamma prior with shape $a$ and \emph{rate} $b$, leading to the update equation 
 \begin{equation}\label{eqn:bayes_em_update_eqn}
\lambdaIter{t+1} =   \frac{N+a-1}{b + \sum_{i=1}^N\sum_{j=1}^{k_i} \frac{1}{\lambdaIter{t}+j} }. 
 \end{equation}
 We see that is has the likelihood EM algorithm as a special case when $a=1$ and $b=0$. In this paper we use these settings for $a$ and $b$ in absence of prior information.  
We used the EM update representation in Equation \ref{eqn:em_used} because this eliminated an additional summation in our Python code and enabled the use of list comprehensions, improving the runtime of the code while maintaining exactly the same values as Equation \ref{eqn:em_used}.
\begin{algorithm}
    \caption{Yule-Simon EM algorithm}
    \label{algo:vys}
    \begin{algorithmic}[1] 
        \Procedure{YS-EM}{$a,b,\mathbf{k}, \lambdaIter{0}$} \Comment{Arguments}
            \While{$\Delta \lambdaIter{t+1} \geq \varepsilon$} \Comment{tolerance $=\varepsilon$}
			\State $\lambdaIter{t+1} = \frac{ N+a-1 }{ b + \sum_{i=1}^N\sum_{j=1}^{k_i}\frac{1}{\lambdaIter{t}+j}}$
			\State $\Delta \lambdaIter{t+1} \gets \vert \lambdaIter{t+1} - \lambdaIter{t}\vert$
            \EndWhile\label{YS-EM endwhile}
            \State \textbf{return} $\lambdaIter{t+1}$ \Comment{Additional diagnostics may be added.}
        \EndProcedure
    \end{algorithmic}
\end{algorithm}
The EM approach has the added benefit that the $\lambda$ update as a function of the previous iteration arises explicitly from the EM process while the derivation of the fixed point algorithm makes an \emph{ad hoc} observation that one could replace the right-hand formula $\lambda$ with that of a previous iteration. In Garcia's derivation, there is no uncertainty quantification based on a standard error, nor a proof of convergence. We next proceed to give standard error estimates and convergence certificates for this EM algorithm.

\section{STANDARD ERRORS AND CONVERGENCE}\label{sec:SEs}

In this section we give standard errors for the EM estimates from the solution to Equation \ref{eqn:em_update_eqn}.
Standard errors are desired quantities to determine if two point estimates differ in a statistically meaningful way. We take two approaches toward the calculation of standard errors for the EM estimates: the Louis formula \cite{louis1982finding} and the Oakes \cite{oakes1999direct} formula. 
Both estimation methods provide the same numerical value. 

\subsection{Standard Error of the EM Estimates}

The general form of the Oakes and Louis equations are given in terms of the Fisher information,
 \begin{equation}\label{eqn:oakes}
\mathcal{I}_O = -\left(\frac{\partial^2Q(\lambda\vert \lambdaIter{t})}{\partial \lambda^2} + \frac{\partial^2Q(\lambda\vert \lambdaIter{t})}{\partial \lambdaIter{t}\partial \lambda}\right)\biggr\rvert_{ \lambdaIter{t} =  \lambda= \lambdaHatEM}, 
 \end{equation} 
 and by the Louis equation, 
 \begin{align}\label{eqn:louis}
\mathcal{I}_L &= E_{\lambda}\left[B(\bf{p},\bf{k},\lambda)\right] \nonumber \\ 
&- E_{\lambda}\left[S(\bf{k}, \bf{p},\lambda)S^T(\bf{k}, \bf{p},\lambda)\right] + S^*(\bf{k},\lambda)S^{*T}(\bf{k},\lambda).
\end{align}
In Equation \ref{eqn:louis} the expression is evaluated at the last iteration of the EM algorithm, making the last term, $S^*(\bf{k}, \lambdaHatEM)$, equal to zero. In Equation \ref{eqn:louis}, $S$ and $S^*$ denote the score functions (partial derivative of the log-likelihood) of the complete data and the observed data respectively. Also, $B(\bf{p},\bf{k},\lambda)$ denotes the second derivative of the complete data log-likelihood with respect to $\lambda$. 

The two expressions given by Oakes and Louis lead to numerically equivalent expressions, as explained by Oakes \cite{oakes1999direct}. However, as Oakes notes, it is much simpler to use Equation \ref{eqn:oakes} than to use Equation \ref{eqn:louis}. 
 From the information equations, the standard errors may be estimated via applying the Cramer-Rao lower bound yielding 
  \begin{equation}\label{eqn:varo_equation}
\mathbb{V}ar_{O}(\lambdaHatEM) \approx \frac{1}{\frac{N}{\left(\lambdaHatEM\right)^2} - \sum_{i=1}^N\sum_{j=1}^{k_i}\frac{1}{\left(j+\lambdaHatEM \right)^2}}.
 \end{equation} 
 The appendix contains the details of the derivations for Equations \ref{eqn:oakes}, \ref{eqn:louis} and \ref{eqn:varo_equation}. 
\subsection{Convergence}
This section draws largely on the exposition in McLachlan and Krishnan \cite{mclachlan2007algorithm} on convergence of the EM algorithm. Note that in this section we use the word convex to mean convex downwards because we are working with maximization operations.
To verify the EM algorithm converges we use a Corollary from McLachlan and Krishnan (pg. 84) which we state here for completeness. 

\textbf{Corollary:} Suppose that $\ell(\lambda)$ is unimodal in a set $\Omega$ with $\lambdaIter{\infty}$ being the only stationary point and $\partial Q(\lambda \vert \lambdaIter{t})/\partial \lambda$ is continuous in $\lambda$ and $\lambdaIter{t}$. Then any EM sequence $\{\lambdaIter{t}\}$ converges to the unique maximizer $\lambdaIter{\infty}$ of $\ell(\lambda)$.  That is, it converges to the unique maximum of $\ell(\lambda)$.

Thus it will suffice for us to show that the log-likelihood is unimodal in the space $\lambda >0$ to show both convergence and that we get the correct estimator. 
To show that $L(\lambda)$ is unimodal we show that $L(\lambda)$ is convex. Recall, that if a function is convex with respect to the argument then any local optimum is a global optimum. We now argue that $L(\lambda)$ is convex over a subset of $\lambda >0$. 

First consider the likelihood, Equation \ref{eqn:ys_w_gammas}, for one $k$. Using the Gamma function recursion $\Gamma(z+1)=z\Gamma(z)$ write the likelihood 
\begin{equation}\label{eqn:factorized}
L(\lambda) = \frac{\lambda\prod_{j=1}^k(\lambda+j)^{-1}\Gamma(k)\Gamma(\lambda+1)}{\Gamma(\lambda+1)}.
\end{equation}
The two gamma functions cancel and taking logarithms we are left with 
\begin{equation}
\ell(\lambda) = log(\lambda) + \sum_{j=1}^k-log(\lambda+j) + log(\Gamma(k)).
\end{equation}
 Taking derivatives twice for $\lambda$ yields 

\begin{equation}\label{eqn:convOfCompleteLikelihood}
\frac{\partial^2\ell(\lambda)}{\partial \lambda^2} = -\frac{1}{\lambda^2} + \sum_{j=1}^k \frac{1}{(\lambda+j)^2}.
\end{equation}
Now we want to determine the interval, denoted $\mathcal{A(\lambda)}$, such that $\frac{\partial^2\ell(\lambda)}{\partial \lambda^2} < 0$. To precisely quantify this interval is a challenging numerical task. For a first pass we will result to a numerical approximation and then seek to refine the approximation. To begin with we see that $\mathcal{A(\lambda)}$ is given by all $\lambda$ that satisfy 
\begin{equation}
  \sum_{j=1}^k \frac{1}{(\lambda+j)^2} < \frac{1}{\lambda^2}.
\end{equation}
Focusing on the left hand side we see that taking $k \to \infty$ will give us a subset of $\mathcal{A(\lambda)}$ where convergence still holds. Additionally, the limiting sum $s(\lambda) = \sum_{j=1}^\infty \frac{1}{(\lambda+j)^2}$ is less than $s(0)=\frac{\pi^2}{6}$, the infinite sum with $\lambda=0$. The numerical value $s(0)$ is known in the mathematical literature as the Basel sum or the Basel problem solution \cite{lagarias2013euler,sandifer2004euler} and Euler was the first to prove the value of the sum. Thus we choose $\mathcal{B(\lambda)} \subset \mathcal{A(\lambda)}$, where $\mathcal{B(\lambda)}$ is the set of $\lambda$ that satisfies $\frac{\pi^2}{6} < \frac{1}{\lambda^2}$ which is $0<\lambda< \frac{\sqrt{6}}{\pi} \approx 0.779696801$. While our simulation experiments indicates that the inequality above holds for a much larger interval of lambda than $0<\lambda< \frac{\sqrt{6}}{\pi}$, without an approach using the complex domain we are currently unable to expand the interval.

While we have shown that the \YS\ EM algorithm converges, the next question to ask is: how \emph{fast} does the EM algorithm converge? We turn to this question in the next subsection.  

\subsection{The Convergence Rate}
The EM algorithm is often said to exhibit slow convergence which is attributed to a sub-linear convergence rate. 
The rate of convergence is determined by the spectral radius, denoted $r$, of the Jacobian of the mapping $M(\lambda)$ given by Equation \ref{eqn:em_used} evaluated at $\lambdaIter{\infty}$, the \emph{self consistency} point. 
Confusingly, different authors define the radius differently, Meng 1994 \cite{meng1994global} defines the radius as $1-r$, making slower values closer to $0$.  Similar to Maclachlan and Krishnan \cite{mclachlan2007algorithm}  we define the radius $r$, so that slower rates are closer to $1$.   

We use the fundamental EM relation of observed information to the ratio of the complete and the unobserved information conditional on observed data. The relationship between these quantities is formalized by 
\begin{equation}
g(k_i \vert \lambda ) = \frac{ f(k_i,p_i \vert \lambda) }{ h(p_i \vert k_i, \lambda) }.
\end{equation}
for $\indices{N}$. Note that here $k_i$ denotes observed data, $p_i$ denotes missing data, and $\lambda$ denotes the parameter(s) to be estimated. Then applying the operator $d^2 log(\cdot) /d \lambda^2$ on both sides yields
\begin{equation}
I_o(\lambda; k_i) = I_c(\lambda;k_i,p_i) - I_m(\lambda, k_i; p_i),
\end{equation}
where the $I_o,I_c,I_m$ indicate the observed, complete, and missing information respectively. Now taking an expectation of both sides with respect to $h(p \vert k_i, \lambda)$ yields 
\begin{equation}
I_o(\lambda; k_i) = \mathcal{I}_c(\lambda;k_i) - \mathcal{I}_m(\lambda;k_i),
\end{equation}
Then, considering the \YS\ case, one can show that $\mathcal{I}_c(\lambda;k_i) = 1/\lambda^2$ and $\mathcal{I}_m(\lambda; k_i) = \sum_{j=1}^{k_i}\frac{1}{(\lambda+j)^2}$. The rate of convergence is 
\begin{equation} \label{eqn:ys_em_rate}
\begin{split}
r \equiv \frac{\mathcal{I}_m(\lambda; p)}{ \mathcal{I}_c(\lambda;k_i,p)} &= \lambda^2(\psi_1(\lambda+k_i+1) - \psi_1(\lambda+1)) \\
 &=\lambda^2\sum_{j=1}^{k_i}\frac{1}{(\lambda+j)^2}, 
\end{split}
\end{equation}
where $\psi_1$ denotes the trigamma function. 
The quantity in Equation \ref{eqn:ys_em_rate} is for a sample size of one. If we consider a sample of $k_i$ with $\lambdaIter{N}$, $N\geq1$, the convergence rate expression becomes
\begin{equation} \label{eqn:ys_em_rate_N}
r \equiv \frac{\lambda^2\sum_{i=1}^N\sum_{j=1}^{k_i}\frac{1}{(\lambda+j)^2}}{N}. 
\end{equation}
While in many models the $M(\cdot)$ mapping is not explicitly available, in the Yule-Simon case we have this update directly via Equation \ref{eqn:em_used}. 
%
%
%
If we consider the sample of observations $k_i$ with $\indices{N}$, evaluated analytically, the Jacobian of the mapping becomes
 \begin{equation}\label{eqn:jacobian_N}
J(\lambda) =  \frac{N \sum_{i=1}^N[\psi_1(\lambda + k_i+1) - \psi (\lambda +1)]}{\left(\sum_{i=1}^N[\psi(\lambda + k_i+1) - \psi(\lambda+1)]\right)^2 }. 
 \end{equation}
 which also evaluates to a scalar at $\lambda = \lambdaIter{EM}$ representing the convergence rate.
 Although a theoretical analysis seems challenging using Equation \ref{eqn:jacobian_N}, the Jacobian can be evaluated numerically to determine $r$ in a range of likely values of $\lambda$.  
\section{EXPERIMENTS} 

We run experiments to empirically evaluate the performance of the EM algorithm, its standard errors calculated via both Louis' and Oakes' equations, as well as the number of iterations to convergence. We also compare the EM algorithm's performance to the Gibbs sampler algorithm's on both synthetically generated data and word counts from five texts.

We repeat the experiments for $N_{rep} = 10,000$ runs. We calculate numerical summaries of the $10,000$ estimates: the mean, median, $95$th percentile and standard deviation.

\subsection{Synthetic Data}

We choose $\lambda$ values that mimic the choices by Garcia and Liesen et al., $\lambda \in \{0.6, 0.8, 1.25, 5, 10\}$. These values of $\lambda$ are also intended to mimic the estimated $\lambda$ values for word frequencies from the five texts. To generate random variates we use the mixture representation in Equation \ref{eqn:latents} to generate $N$ \YS\ random variates ($k_i$).  We chose a sample size $N \in \{50, 500, 5000\}$ to evaluate the algorithm in both small and large sample size settings. 

The standard errors and point estimates are close to-if not the same as-the numbers from the Gibbs sampler. However, going the EM route is quicker, and requires fewer \emph{ad hoc} choices, such as the burn-in and thinning rates needed for the Gibbs sampler.

The steps to generate a sample of size $N$ of synthetic data using the mixture distribution representation are:

\begin{enumerate}
\item Set starting random seed and keep track for reproducibility.
\item Generate a random value $p_i$ between zero and one from a $Beta(\lambda,1)$ distribution.
\item Generate a random number $k_i$ from a Geometric distribution with probability of success $p_i$ equal to the value at the previous step. 
\item Repeat steps two and three $N$ times.
\end{enumerate}

We also replicate a modified Polya urn data generating process described in Garcia and verify the similarity with the samples from the mixture data generating process. However, in a Polya urn data generating process $\lambda$ cannot take values less than one. We see $\lambdaIter{t} <1$ in the text application and therefore are interested in generating synthetic data with $\lambda$ in the full range, $(0,1)$ as well as $\lambda>1$, hence we use the mixture representation data generating process.

We then implement the EM algorithm to get a point estimate for $\lambda$, a Louis standard error, and an Oakes standard error. We track the actual number of iterations to convergence of the EM algorithm, calculate the empirical rate of convergence path, and evaluate numerically the convergence rate given by Equation \ref{eqn:ys_em_rate} where $\hat{\lambda}^{(t)}$ is used to evaluate the formula. 

For comparison purposes, we implement the Gibbs sampler described by Liesen et al. to sample from the posterior of $\lambda$ and calculate a point estimate (the sample mean) and a standard error of the estimate (the sample standard deviation). We generate 8,000 $\lambda$ values via the Gibbs sampler and use a burn in of 500 and no thinning. Similarly to Liesen et al. we use a gamma prior with $a=0.05$ and $b=0.25$. We evaluate different gamma hyperparameters without any significant changes to the Gibbs point estimates. For example $a=b=1$ and $a=0$ and $b=1$ both lead to the same EM estimate of 0.618 and standard error of 0.0053 for the distribution of word frequencies in the text Ulysses. 

We experimented with other values of the Gibbs sampler hyperparameters (for example 50,000 samples). All settings resulted in the same parameter estimates up to three significant digits. We use 8,000 samples to reduce runtime while maintaining numeric precision in the Gibbs estimates. Furthermore, an analysis of the autocorrelation on a sequence of Gibbs runs for all lags up to 100 indicated all autocorrelations were less than $5\%$ and non-significant, hence we do not thin.
 
Figure 2 illustrates the range of median EM (dash line) and Gibbs estimates and standard errors for all considered initial $\lambda$ values and sample sizes $50$ and $500$. 

\begin{figure}[h]
\centering
\vspace{.2in}
\includegraphics[height=3.2in, width=3.2in, keepaspectratio]{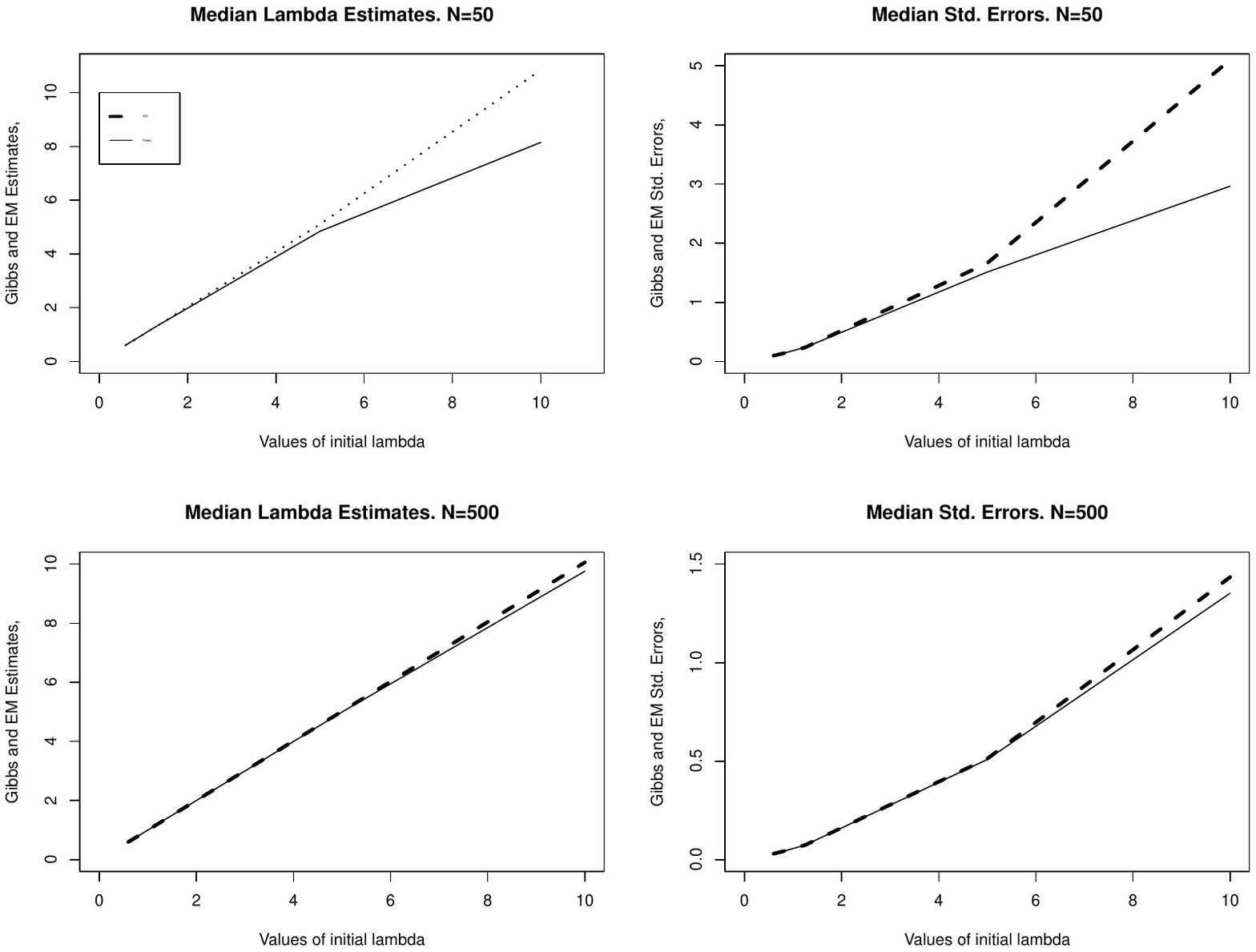}
\vspace{.2in}
\caption{Median EM and Gibbs Lambda Estimates and Associated Standard Error Estimates.}
\label{fig:med}
\end{figure}

For $\lambda=0.6$ the EM and Gibbs estimates coincide in the three considered sample sizes and for either numerical summary (mean, median and $95$th percentile). The standard errors coincide as well up to at least three decimals. The EM algorithm converges in under nine iterations. For larger sample sizes, the standard errors are lower (for example 0.0095 in both Gibbs and EM algorithms on average for $N=5000$ vs 0.0968 for $N=50$). The cases with $\lambda=0.8$ and $\lambda=1.25$ display very similar patterns to the case with $\lambda=0.6$. 

For higher values of lambda (five or ten) values of EM and Gibbs estimates start diverging in small sample sizes. The average standard error of the estimate for Gibbs sampler is significantly lower ($50\%$) than the average standard errors for the EM, although the median standard errors are very similar. For cases with small sample size and high $\lambda$ there are a few experiment runs where the EM takes a large number of iterations to converge and achieves poor estimates with large standard errors.  For example, for the case with initial $\lambda = 5$ the $95$th percentile of the $\lambda$ estimate from 10,000 experimental runs is 10.8 for the EM and 8.1 for Gibbs, with a standard error of five (EM) versus 2.93 (Gibbs). The EM algorithm converges in approximately 50 iterations for the initial $\lambda = 5$ case. The estimates for EM and Gibbs become very close for $N=500$ and match up to three decimals for a sample size of 5,000. 

Overall, with larger sample sizes we see smaller Louis/Oakes standard errors for the EM algorithm.


\subsection{Application to Text}

We next use word frequencies from five texts to evaluate the EM algorithm and compare it to the Gibbs sampler. These texts were also used in Garcia and Leisen et al., and we obtained them from the Gutenberg website. The numerical results will vary with the choice of preprocessing. In contrast with Garcia, we chose to remove the header and footer of the Gutenberg text files. These headers and footers contain a boiler plate description of the terms of use. The language used was mostly specific to the legal profession and does not bear an indication on the original author's choice of words. 

In Table \ref{tab:text1} and Table \ref{tab:text2} we see the estimates for five texts, Ulysses ($N=29,216$ unique words), War and Peace ($N=17,557$), Don Quixote ($N=14,622$), Moby-Dick ($N=16,861$), and Les Miserables ($N=23,451$). 

\begin{table}[!htbp] \centering 
  \caption{Text Application Results}
  \label{tab:text1} 
\scriptsize
\begin{tabular}{@{\extracolsep{0pt}} D{.}{.}{-4} D{.}{.}{-4} D{.}{.}{-4} D{.}{.}{-4} D{.}{.}{-4} D{.}{.}{-4} } 
\multicolumn{1}{l}{\bf{ESTIMATE}}&\multicolumn{1}{c}{\bf{U}} & \multicolumn{1}{c}{\bf{WP}} \\ 
\hline \\[-1.8ex] 
\multicolumn{1}{l}{EM $\lambda$} & 1.0780 & 0.6181 \\ 
\multicolumn{1}{l}{EM iterations} &  9 & 8 \\ 
\multicolumn{1}{l}{Gibbs $\lambda$} & 1.0781 & 0.6181 \\ 
\multicolumn{1}{l}{Gibbs Std Err} &  0.0079 & 0.0053 \\ 
\multicolumn{1}{l}{Louis/Oakes Std Err} & 0.0080 & 0.0053 \\ 
\end{tabular} 
\end{table}

\begin{table}[!htbp] \centering 
  \caption{Text Application Results (cont.)} 
  \label{tab:text2} 
\scriptsize
\begin{tabular}{@{\extracolsep{0pt}} D{.}{.}{-4} D{.}{.}{-4} D{.}{.}{-4} D{.}{.}{-4} D{.}{.}{-4} D{.}{.}{-4} } 
\multicolumn{1}{l}{\bf{ESTIMATE}} & \multicolumn{1}{c}{\bf{LM}} & \multicolumn{1}{c}{\bf{MD}} & \multicolumn{1}{c}{\bf{DQ}}  \\ 
\hline \\[-1.8ex] 
\multicolumn{1}{l}{EM $\lambda$} & 0.7028 & 0.8679 & 0.6696  \\ 
\multicolumn{1}{l}{EM iterations} & 9 & 10 & 9 \\ 
\multicolumn{1}{l}{Gibbs $\lambda$} & 0.7028 & 0.8682 & 0.6696 \\ 
\multicolumn{1}{l}{Gibbs Std Err} & 0.0053 & 0.0080 & 0.0064 \\ 
\multicolumn{1}{l}{Louis/Oakes Std Err} & 0.0053 & 0.0081 & 0.0064 \\ 
\end{tabular} 
\end{table} 

where U stands for Ulysses, WP stands for War and Peace, MD for Moby-Dick, LM for Les Miserables and DQ stands for Don Quixote.

After rounding to 4 significant digits, the point estimates from both EM and Gibbs procedures are identical for all five texts.



\subsection{Empirical Convergence Rates}

\begin{figure}[h]
\centering
\vspace{.2in}
\includegraphics[height=3.5in, width=3.5in, keepaspectratio]{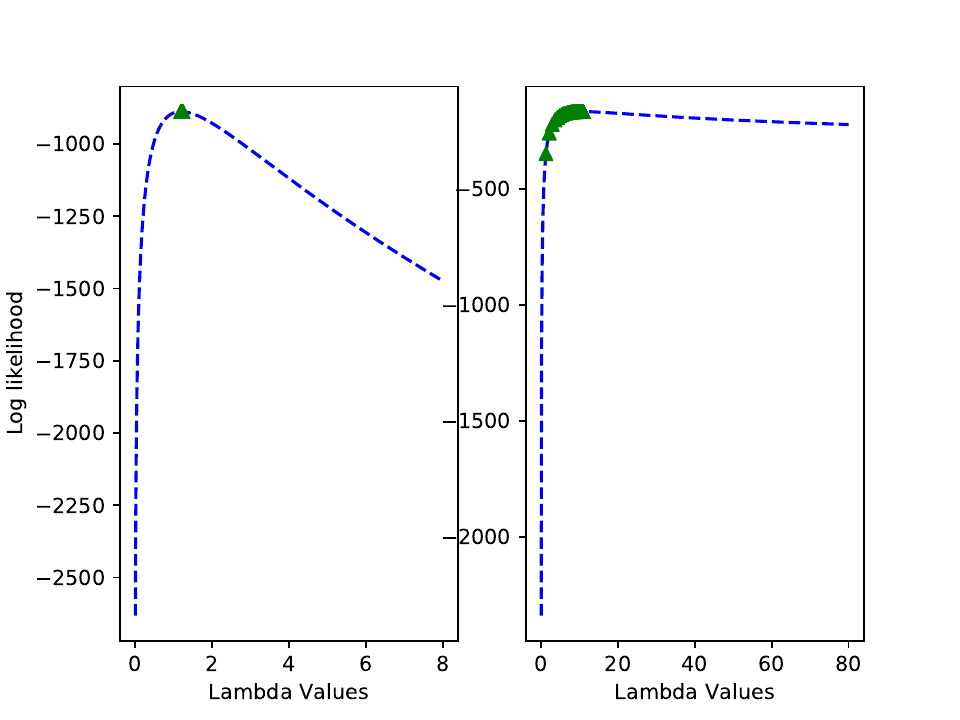}
\vspace{.2in}
\caption{Log-likelihood for synthetic data with true $\lambda = 1.25$ (left figure) and $\lambda = 10$ (right figure). $N=500$. The triangles indicate successive iterations of $\lambdaIter{(t)}$ values on the horizontal axis and log-likelihood values on the vertical axis.}
\label{fig:loglhood}
\end{figure}

In all the experiments the EM algorithm required less than 10 iterations for small values of $\lambda$. With the same starting values and tolerance, for larger initial $\lambda$ values of 5 or 10, the EM required more iterations (50 to 100). The primary reason for a larger number of iterations is a flat log-likelihood near the optimal point, as seen in Figure \ref{fig:loglhood}. 

In addition to keeping track of the actual number of iterations required to converge, we also calculate empirical convergence rate paths and numerically evaluated the theoretical convergence rate formula. 

For the text data, the empirical rate of convergence is  
\begin{equation}\label{eqn:emp_conv_rate}
r^{(t+1)} = \frac{\lambdaIter{t+1} - \lambdaIter{t} }{\lambdaIter{t} - \lambdaIter{t - 1}}.
\end{equation}

 For synthetically generated data we calculate the empirical convergence rate as
\begin{equation}\label{eqn:sim_conv_rate}
r^{(t+1)} = \frac{\lambdaIter{t} - \lambdaIter{\infty}}{\lambdaIter{t-1} - \lambdaIter{\infty}},
\end{equation}
at each iteration. 

\begin{figure}[h]
\centering
\vspace{.2in}
\includegraphics[height=3.2in, width=3.2in, keepaspectratio]{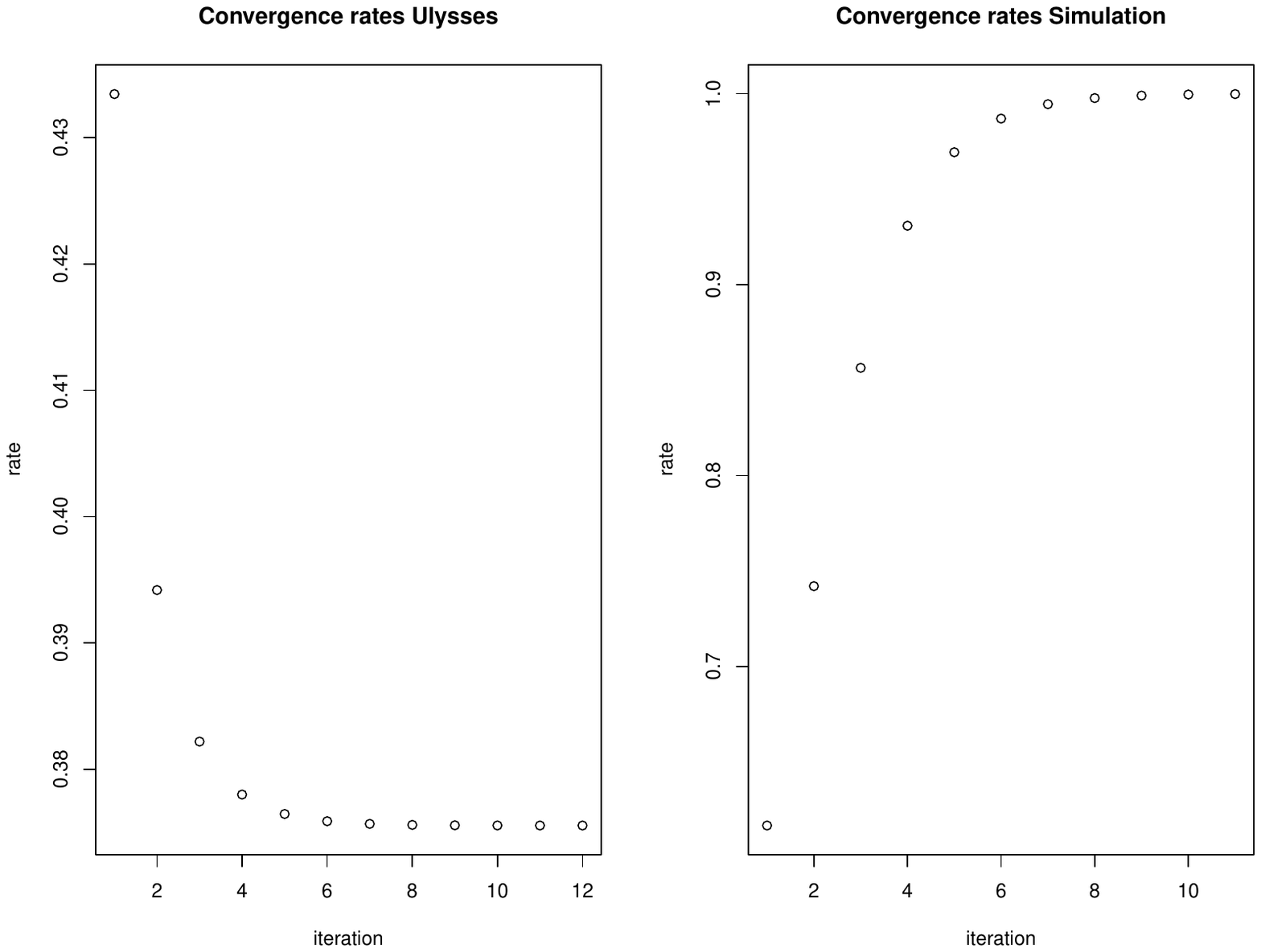}
\vspace{.2in}
\caption{Empirical Convergence paths for Ulysses (left) and synthetic data (right) with true $\lambdaIter{\infty}=1.25$.}
\label{fig:conv_paths}
\end{figure}

We illustrate the empirical convergence rate for the Ulysses text using Equation \ref{eqn:emp_conv_rate} and for synthetic data using Equation \ref{eqn:sim_conv_rate}, with true $\lambdaIter{\infty} = 1.25$ in Figure \ref{fig:conv_paths}.

Next we numerically evaluated the theoretical convergence rate $r$ given by Equation \ref{eqn:ys_em_rate_N} using synthetically generated data. The theory posed that we have linear convergence for $0<r\ll1$ and sub-linear convergence for values close to $r\cong1$. For data generated synthetically with $\lambda = 1.1$ in a sample size of $N=500$, $r$ is approximately $0.38$. This is consistent with the empirical convergence results. In a case with larger $\lambda$, the calculated theoretical convergence rate $r$ for a generated sample of $N=500$ data points is approximately 0.89 (slower), which is again consistent with the observed empirical convergence.

\subsection{A Note on Starting Values for $\lambda$ in the EM Algorithm}

The EM algorithm starts with an initial value denoted $\lambda^{(0)}$. However, an experiment ranging $\lambda^{(0)}$ in the EM algorithm reveals that there is not a significant difference in either speed of convergence or the final estimate. For the Ulysses text, the EM algorithm with starting values $\lambda^{(0)} = \{ 0, 0.6, 1.1, 1.25, 2\}$, led to the same EM estimate up to three decimals ($0.618$) in seven iterations. Only when $\lambda^{(0)} =0.6$ the algorithm converged in four iterations because the start was very close to the EM estimate. In the synthetic data experiments we experimented with $\lambda^{(0)}=\{0,1.1\}$ with very similar results in both cases. Some other choices for starting values can be for example the mode, $\lambda^{(0)} =1$ or, when we expect $\lambda$ to be above one (considering that the first moment is only defined for $\lambda > 1$), the method of moments estimator, $\lambdaHatM = \frac{\bar{k}}{\bar{k} - 1}$, where $\bar{k}$ denotes the arithmetic mean of the observed counts $k_i$. 

\section{CONCLUSION}

We presented an EM algorithm for the estimation of $\lambda$ in a \YS\ distribution. We concluded that both the fixed point and the EM algorithm estimates will follow the same solution path. We constructed standard errors for the resulting EM estimate that did not exist in the literature. We also compared the estimates from the EM algorithm with the estimates from a Gibbs sampler of the posterior for $\lambda$. We were able to determine whether these estimates differed substantially using the derived standard error for the EM estimate. We also proved the convergence of the EM algorithm for the \YS\ distribution, a result not previously shown. Moreover, we quantified the rate of convergence of the EM algorithm and determined regimes of linear and sub-linear convergence. We evaluated the estimation approach in experiments with both synthetic data and written text frequencies. We conclude that the EM algorithm provides a convergent and statistically sound method to estimate the $\lambda$ parameter in a \YS\ process.

\bibliographystyle{plain}   
\bibliography{notes_on_yule_simon} 

\appendix\newpage\markboth{Appendix}{Appendix}
\renewcommand{\thesection}{\Alph{section}}
\numberwithin{equation}{section}
\section{APPENDIX}

\subsection*{Beta Entropy Moment Formulae}
We may use symmetry of the beta random variable to switch from $p_i$ to $(1-p_i)$ to get the formula for both expectations using the formulae  
\begin{equation} \label{eqn:exp}
\begin{split}
 &\E[log(p_i)] = \psi(\lambdaIter{t} + 1) - \psi(\lambdaIter{t} + k_i + 1),\\
 &\E[log(1 - p_i)]= \psi(k_i) - \psi(\lambdaIter{t} + k_i + 1),
 \end{split}
 \end{equation}
where $\psi$ denotes the digamma function, the derivative of the natural log of the gamma function. 

Also, if $p$ is a $Beta(\alpha,\beta)$ random variable then
\begin{equation}\label{eqn:exp_log_beta_z}
\mathbb{E}(log(p)^z) = \psi_{z-1}(\alpha) - \psi_{z-1}(\alpha+\beta)
\end{equation}
To derive Equation \ref{eqn:exp_log_beta_z} we consider the \emph{unnormalized} density function for a beta random variable
\begin{align}
&\int_0^1\left[log(p)\right]^zp^\alpha\frac{(1-p)^{\beta-1}}{p} dp \\
&= \int_0^1\left[\frac{d^z}{d\alpha^z}\right]p^\alpha\frac{(1-p)^{\beta-1}}{p} dp\nonumber \\ \nonumber
&=\left[\frac{d^z}{d\alpha^z}\right] \int_0^1p^{\alpha-1}(1-p)^{\beta-1} dp
\end{align}
Now on the right hand side of the equation above we multiply back through by the normalizing constant $B(\alpha, \beta) =\frac{\Gamma(\alpha)\Gamma(\beta)}{\Gamma(\alpha+\beta)}$ to get an equation that is properly interpreted as an expectation. 
Now setting $z=1$ and noting that the equation is
\begin{align*}
\frac{\frac{d}{d\alpha}B(\alpha, \beta)}{B(\alpha, \beta)} 
&= \frac{d}{d\alpha} log(B(\alpha, \beta))\\
&=\frac{d}{d\alpha} \left( log(\Gamma(\alpha)) + log(\Gamma(\beta))- log(\Gamma(\alpha+\beta)) \right)\\
&=\psi(\alpha) - \psi(\alpha+\beta),
\end{align*}
the same equation as quoted for the entropy of a Beta random variable. A similar argument for arbitrary non-negative integer valued $z$ will yield Equation \ref{eqn:exp_log_beta_z}. Finally, note by symmetry considerations if the expectation is $\mathbb{E}([log(1-p)]^z)$ you may simply switch the roles of $\alpha$ and $\beta$ in Equation \ref{eqn:exp_log_beta_z}. 

\subsection*{Oakes Standard Error Derivation for \YS\ Model}

The Oakes formula \cite{oakes1999direct} is given by Equation \ref{eqn:oakes} in the manuscript. Taking the second derivative of the $Q(\lambda \vert \lambdaIter{t})$ given by Equation \ref{eqn:q_func} is equivalent to taking the first derivative with respect to ${\lambda}$ of Equation \ref{eqn:partial_q}, 
\begin{equation}\label{eqn:o_derivation}
\begin{split}
\frac{\partial^2Q(\lambda\vert \lambdaIter{t})}{\partial \lambda^2} = -\frac{N}{\lambda^2}.
\end{split}
\end{equation}

Now the second term of Equation \ref{eqn:oakes} arises from taking the first partial with respect to ${\lambdaIter{t}}$ from Equation \ref{eqn:partial_q}. Putting these two equations together we obtain,

\begin{equation}\label{eqn:o1}
\frac{\partial Q(\lambda\vert \lambdaIter{t})}{\partial \lambdaIter{t}} = N\psi_1 ( \lambdaIter{t} +1) - \lambda\sum_{i=1}^N\psi_1( \lambdaIter{t} +1 + k_i) .
\end{equation}

After taking another partial derivative we get an expression with trigamma functions denoted $\psi_1$, the derivative of the digamma function. Using the trigamma recurrence 

\begin{equation}\label{eqn:trigamma}
\psi_1(z+1) = \psi_1(z) - \frac{1}{z^2}
\end{equation}

 Equation \ref{eqn:o1} becomes

\begin{equation}\label{eqn:o2}
\begin{split}
\frac{\partial^2Q(\lambda\vert \lambdaIter{t})}{\partial \lambda\partial \lambdaIter{t}} = N\psi_1 ( \lambdaIter{t} +1) - N\psi_1 ( \lambdaIter{t} +1) +\\
 \sum_{i=1}^N\sum_{j=1}^{k_i} \frac{1}{{\left(\lambdaIter{t} + j\right)}^2} 
\end{split}
\end{equation}

Then \ref{eqn:oakes} becomes

\begin{equation}\label{eqn:o3}
\mathcal{I}_O = \frac{N}{\lambda^2} - \sum_{i=1}^N\sum_{j=1}^{k_i} \frac{1}{{\left(\lambda^{(t)} + j\right)}^2}
\end{equation}

By replacing $\lambda$ and $\lambdaIter{t}$ with the EM estimate at convergence and taking the inverse of Equation \ref{eqn:o3} we obtain 
Equation \ref{eqn:varo_equation}.

\subsection*{Louis Standard Error Derivation for \YS\ Model}

The Louis formula \cite{louis1982finding} for the standard error is given by Equation \ref{eqn:louis}. 
We will use the notation used by Louis to simplify the derivations.  The gradient of the complete data log-likelihood is $S(X,\lambda)$. We denote with $B(X, \lambda)$ the minus second derivative of $l$, the complete data log-likelihood. The $S^*$ is the gradient of $Q$ and then evaluated on the last iteration of the EM, it becomes zero.

In the \YS\ case 

\begin{equation}
S^* = \frac{\partial Q(\lambda \vert \lambdaIter{t})}{\lambda} = \frac{N}{\lambda}+\sum_{i=1}^N (\psi(\lambdaIter{t} + 1) -\psi(\lambdaIter{t} + 1 + k_i))
\end{equation}
which we evaluate at the last EM iteration step, replacing $\lambda$ by the EM estimate and $\lambdaIter{t}$ with the estimate at one step before EM convergence. 

In \YS\ case we only have one parameter, $\lambda$, so B, S and $S^*$ are scalars rather than vector valued quantities which may be the case in Louis' formula. Furthermore, in this case the score vector is also a scalar $S^T = S$ and the marginal score $S^* = S^{*T}=0$ when evaluated at the EM estimate.

Next we turn our attention to calculating the first and second terms of the Louis formula. The quantity $B$ is easily found to be

\begin{equation}
B = \sum_{i=1}^N B_i = \frac{N}{\lambda^2}
\end{equation}
where
\begin{equation}
B_i = \frac{\partial S_i}{\partial \lambda} = \frac{1}{\lambda ^2}
\end{equation}
and the score function for one datapoint $i$ is
\begin{equation}
\begin{split}
S_i &= \frac{\partial{l_i}}{\partial\lambda} \\
&= log(p_i) + \frac{1}{\lambda}.
\end{split}
\end{equation}  

In the case of the second term in Equation \ref{eqn:louis}, assuming independence of $X_i$
we can derive

\begin{align}
\mathbb{E}(SS^T) &= \mathbb{E}(S^2) \\
&= \mathbb{E}([\sum_{i=1}^N S_i]^2) \\
 &=\sum_{i=1}^N \mathbb{E}(S_i ^2) + 2\sum_{\substack{i,j=1 \\ i< j}}^N \mathbb{E}(S_i)\mathbb{E}(S_j)
\end{align}

So according to Equation 3.2 in \cite{louis1982finding}

\begin{equation}\label{eqn:final_l}
\begin{split}
I_L(\lambda) = \frac{N}{\lambda ^2}-\sum_{i=1}^N \mathbb{E}(S_i ^2) - 2\sum_{\substack{i,j=1 \\ i< j}}^N \mathbb{E}(S_i)\mathbb{E}(S_j).
\end{split}
\end{equation}

We first need to calculate

\begin{equation}
\mathbb{E}(S_i ^2) = \mathbb{E}[(log(p_i))^2] + \frac{1}{\lambda ^2} + 2\frac{\mathbb{E}(log(p_i))}{\lambda}
\end{equation}

To get expected values we use the first expression in Equation \ref{eqn:exp} and the following known result
\begin{equation}\label{eqn:var}
Var(X) = \mathbb{E}(X^2) - (\mathbb{E}(X))^2
\end{equation}

Putting it all together
\begin{equation}\label{eqn:si_sq}
\begin{split}
\mathbb{E}(S_i ^2) &= \sum_{i=1}^N[\psi_1(\lambda + 1) - \psi_1(\lambda + k_i + 1)] \\
& +\sum_{i=1}^N[[\psi(\lambda + 1) - \psi(\lambda + k_i + 1)]^2] + \frac{N}{\lambda ^2}\\
 &+ \frac{2}{\lambda} \sum_{i=1}^N(\psi(\lambda + 1) - \psi(\lambda + k_i + 1))
 \end{split}
\end{equation}

The second part of the second term depends on the calculation of 
\begin{equation}
\begin{split}
\mathbb{E}(S_i) &= \frac{1}{\lambda} + \mathbb{E}(log(p_i)) \\
&= \frac{1}{\lambda} + (\psi(\lambda + 1) - \psi(\lambda + k_i + 1))
\end{split}
\end{equation}

One can use the recurrence for the digamma and trigamma functions to eliminate the need for using special functions. From the digamma recurrence in Equation \ref{eqn:digamma} we have the simplification

\begin{equation}
\psi(\lambda + k_i + 1) - \psi(\lambda + 1) = \sum_{t=1}^{k_i} \frac{1}{(\lambda + t)}.
\end{equation}

From the trigamma recurrence in equation \ref{eqn:trigamma} we get the simplification

\begin{equation}
\psi_1(\lambda + 1) - \psi_1(\lambda + k_i + 1) = \sum_{t=1}^{k_i} \frac{1}{(\lambda + t)^2}.
\end{equation}
Note that we experimented in Python with the finite sum representation and a direct call to an implementation of the digamma and trigamma functions. We chose the latter because the code was simplified and runtime improved from our judicious use of list comprehensions.
Then 
\begin{equation}
\begin{split}
\mathbb{E}(S_i ^2) & = \frac{N}{\lambda^2} + \sum_{i=1}^N \sum_{t=1}^{k_i} \frac{1}{(\lambda + t)^2} \\
& +\sum_{i=1}^N \left[\sum_{t=1}^{k_i} \frac{1}{(\lambda + t)}\right]^2  \\
& - \frac{2}{\lambda} \sum_{i=1}^N \sum_{t=1}^{k_i} \frac{1}{(\lambda + t)}
\end{split}
\end{equation}

In the same vein we can simplify the term
\begin{equation}
\begin{split}
&2\sum_{\substack{i,j=1 \\ i< j}}^N \mathbb{E}(S_i)\mathbb{E}(S_j) =\\
&2 \sum_{\substack{i,j=1 \\ i< j}}^N(\frac{1}{\lambda} - \sum_{t=1}^{k_i} \frac{1}{(\lambda + t)})(\frac{1}{\lambda} - \sum_{z=1}^{k_i} \frac{1}{(\lambda + z)})
\end{split}
\end{equation}

Then we calculate $I_Y$ using Equation \ref{eqn:final_l} and replacing $\lambda$ everywhere with the EM estimate

\begin{equation}
Var_L = \frac{1}{I_L}.
\end{equation}

The Louis standard error of the EM estimate is the square root of $Var_L$ and it is equivalent to the Oakes standard error. Calculating Oakes is easier and both formulas have simplifications in the \YS\ case that do not depend on special functions.

\end{document}